\documentstyle[twocolumn,prl,aps,epsfig]{revtex}
\begin{document}
\draft
\twocolumn[\hsize\textwidth\columnwidth\hsize\csname
@twocolumnfalse\endcsname

\title{Persistence and Quiescence of Seismicity on Fault Systems}
\author{Matt W. Lee $^1$, Didier Sornette $^{2}$ and Leon Knopoff $^1$}
\address{$^1$ Department of Physics and Astronomy and Institute of
Geophysics and Planetary Physics\\
University of California, Los Angeles, CA 90095}
\address{$^2$ Department of Earth and Space Sciences and Institute of
Geophysics and Planetary Physics\\
University of California, Los Angeles, California 90095-1567\\ and
Laboratoire de Physique de
la Mati\`{e}re Condens\'{e}e\\ CNRS and Universit\'{e} de
Nice-Sophia Antipolis, Parc Valrose, 06108 Nice, France}
\date{\today}
\maketitle
\begin{abstract}
We study the statistics of simulated earthquakes in a quasistatic
model of two parallel heterogeneous faults within a slowly driven
elastic tectonic plate. The probability that one fault remains dormant
while the other is active for a time $\Delta t$ following the previous
activity shift is proportional to $\Delta t^{-(1+x)}$, a result that is robust
in the presence of annealed noise and strength weakening.
A mean field theory
accounts for the observed dependence of the persistence exponent $x$ as a
function of
heterogeneity and distance between faults. These results continue to hold if
the number of competing faults is increased.
\end{abstract}
]
 
Each of the largest earthquakes in recent decades in Southern
California has occurred on a different fault of a
complex network. In contrast, most efforts at modeling seismicity
and its complexity have been
directed at studies of self-organization on a single fault.
We show quantitatively that coupling between
inhomogeneous faults leads to persistent
complementary alternation of quiescence and activity between the faults.
This is related to the persistence phenomenon
discovered in a large variety of
systems \cite{persistence},
which specifies how long a relaxing dynamical system remains
in a neighborhood of its initial configuration. Our persistence exponent
is found to vary as a function of
heterogeneity and distance between faults, thus defining a novel
universality class.
 
Competition between faults has been previously observed qualitatively
in a quasi-static self-organizing earthquake-fault model
\cite{Milten},
and in a dynamic model of the seismicity on two coupled parallel faults
\cite{Knopoff}.
Our present model is a hybrid of these models of interactive faults.
As in \cite{Milten}, the driving stress
is applied and the  stress upon fracture is redistributed
quasistatically through an elastic medium of large extent;
as in \cite{Knopoff}, ruptures can occur on either of two predefined
coupled, parallel
faults, each with variable friction.
Our model is a crack version \cite{Vanneste} of the 2-D quasistatic earthquake
model of a tectonic plate with long range elastic
forces \cite{Milten}. A thin tectonic plate located in the $x$-$y$ plane
undergoes antiplane scalar shear deformation along the $z$-axis.
The plate is discretized in $L\times L$ plaquettes of lattice size $a$
that are oriented at $45$ degrees with respect to the edges;
elastic bonds connect the nodes of the lattice. We impose a slow, uniform
velocity $V_z$ between the two opposite edges
parallel to the $x$-axis.  The system is periodic in the $x$-direction. We
imbed two linear faults parallel to the $x$-axis;
we can consider more than two faults. The faults have finite, spatially
variable, rupture strengths; the other parts of
the system are assumed to be much stronger and never rupture. To ensure
isotropy, we rupture nodes at
the intersections of bonds, rather than individual bonds. The system is
loaded until the total stress on one fault
node reaches a predefined threshold. At this instant, loading is suspended and
the corresponding bonds rupture with a
complete relaxation of stress followed by a redistribution
of stress within the lattice. If additional nodes
become critical, the fracture will continue to
grow until no further ruptures occur; this defines the end of an event. At
this time, the broken bonds are restored to
full strength, and a constant, externally derived, irreversible slip is
applied to each of the broken bonds to correspond
to the force drop associated with the event. Loading now resumes and the
process continues. All bonds have the same elastic
constant in all simulations.
 
In contrast to earlier simulations \cite{Milten,Vanneste}, there is no
spatially
variable quenched disorder on the faults
and we instead introduce annealing of the friction by randomizing the
thresholds
of ruptured nodes after each rupture
to the values $B R(x)$ where $R(x)$ is uniformly distributed in the interval
$[1-r,1+r]$, and $B$ is the average value of the
force rupture threshold. Physically, the randomization accounts for the fact
that a slip associated with an earthquake
brings different asperities in contact that were previously separated,
thereby changing the local sliding threshold.
 
Crack models of this type, with relatively homogeneous
fracture thresholds, often display runaways \cite{Vanneste,Xu},
which are finite-size events that span the entire length of a fault, and
which reset the stress field everywhere along the fault to
zero. Runaways arise because the stress enhancement $\sigma\sqrt{L/a}$
at the crack tips of
fractures of size $L$ can overcome any
heterogeneity bounded by $B(1+r)$ for a sufficiently large crack. Real
earthquake fractures have other length scales, and thus do
not have force concentrations that become excessively large due to
scaling by the length of the fracture. We are concerned that our finding of
the persistent complementary alternation of quiescence and activity with
a power law distribution of time scales might result from
the strong overprint of the stress correlations induced by
runaways, and thus might be dominated by finite size effects. We have thus
extended our model by introducing strength
weakening\,: as soon as the force on a bond reaches
$50\%$ of its initial breaking strength, the force
threshold for rupture of a bond is assumed to decrease linearly with
time  with a decay rate that is twice the loading
rate. The results below do not depend on the specific values of these
two parameters as long as strength weakening
is present at a rate larger than but comparable to the tectonic
loading rate, and the threshold for the onset of weakening is
sufficiently small. Strength weakening simulates the effect of stress
corrosion in fault zones. Weakening
causes ruptures to occur earlier, and thus the energy
available for prolonged fracture growth is reduced. In our simulations,
we have not observed runaways in these conditions.
 
\begin{figure}
\epsfig{file=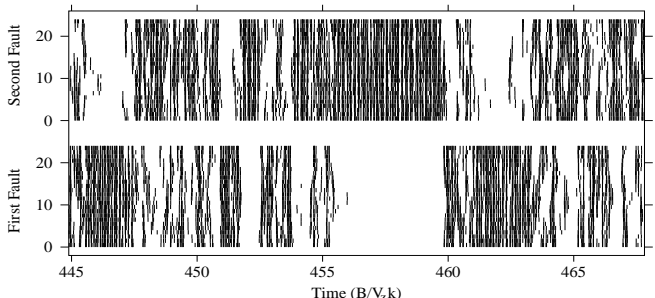, width=8.6cm,height=4cm}
\caption{Space-time history of fault activation in a system of
$24 \times 24$ nodes with two faults parallel to the x-axis.
We have simulated systems up to $96 \times 96$ with similar results.}
\end{figure}
 
Due to the conservative nature of elastic deformation,
at least one fault must be active
to release the average tectonic loading \cite{Milten}. In the case of two
competing faults,
a spontaneous symmetry-breaking occurs, which is an example
of a noise-induced transition \cite{noiseinduced}\,:
there is a well-defined threshold $r_c$ for the amplitude of the annealed
noise below
which all rupture events take place on only one of the two faults and there
is no activity on the other fault. The two faults are not
decoupled, otherwise they would both be independently active of each
other; only their stress fluctuations are decoupled. $r_c$
increases with the distance between the two faults. Fig.~1
is an example of the space-time representation of the
activity on two competing faults for $r = 0.45 > r_c$.  Above $r_c$, the
symmetry is restored and
the rupture events occur on average equally between the two
faults.
 
Shortly after the time that
a quiescent fault
becomes active, activity on the other may cease along its entire length.
The onset of quiescence does not take place
instantly, but does so in
a finite time after the complementary flip to activity on the other
fault.  Activity on the quiescent fault is initiated locally at first,
and then
fractures may appear progressively along its entire length.
Occasionally, ruptures on a quiescent fault start to spread,
but the effort does not lead to full-fledged activity, as at time 455 on
the first fault of the pair.
Once an initial seed fracture has been planted on the quiescent fault,
it can initiate rupture on its neighbors because of the local high
stresses at the edge of the fracture.  Opposite these sites, fractures
on the active fault are extinguished because they are now in the
stress shadow of the newly occurred fractures on the quiescent member.
The two faults are now in competition\,: fractures on
the formerly quiescent fault attempt to spread and
fractures on the formerly totally active fault, attempt to spread into
the ``hole'' of inactivity at the sites complementary to the `quiescent'
fault.  When the two faults are both preoccupied with activity
partially distributed along their lengths, it is impossible to
decide which had been the active and which the quiescent fault.  The
formerly quiescent fault may return to a state of quiescence or it may
become completely active.  Thus there is a finite interval of time between
the triggering of activity on the quiescent fault and the extinction
of activity on an active fault in the case of a `flip'. In many
cases, there is an outburst of activity, but no flip takes place.

 \begin{figure}
\epsfig{file=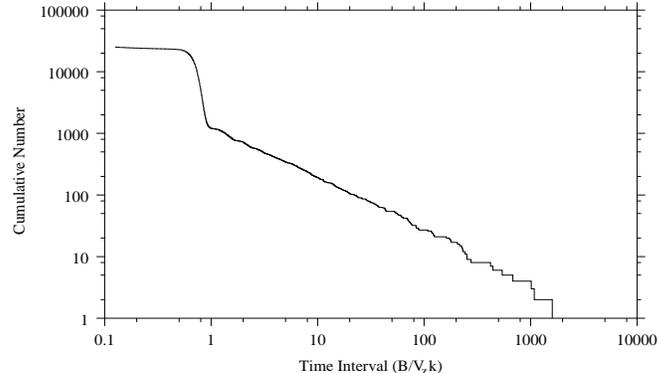, width=8.6cm,height=5cm}
\caption{Cumulative distribution of quiescent time intervals measured in units
of $B/V_z k$
along one of the faults. The distribution for the other fault is similar.}
\end{figure}

To define fault activity, we take a moving time average of
the total slip in all rupture events
on a single fault.  A fault is defined to be active when the window-average
slip is larger than a given threshold. Fig.~2
shows the cumulative
distribution of quiescent time intervals $\Delta t$ along one of the
faults. The results are robust with
respect to the threshold and size of the window, for large enough
$\Delta t$ in comparison with the width of the window. For small
$\Delta t$,
the distribution is that of the time intervals between two consecutive
events on the {\it same} fault, which are scaled by
the tectonic loading time $B/V_zk$, where $k$ is the elastic constant.
Greater than this time, the distribution is exclusively that
for intervals of flipping of activity between the two faults. The differential
distribution of the quiescent intervals $\Delta t$ is a power law
$$
P(\Delta t) \sim 1/\Delta t^{1+x(r, d)}~,   \eqno(1)
$$
for all $r > r_c(d)$.
The larger the value of $r$, the greater the number of flips per unit
time. The exponent $x$ thus increases with $r$ (see Fig.3),
decreases with increasing fault separation $d$ and is independent of
the level of the relaxation threshold and the relaxation
rate within the error bars.

  \begin{figure}
\epsfig{file=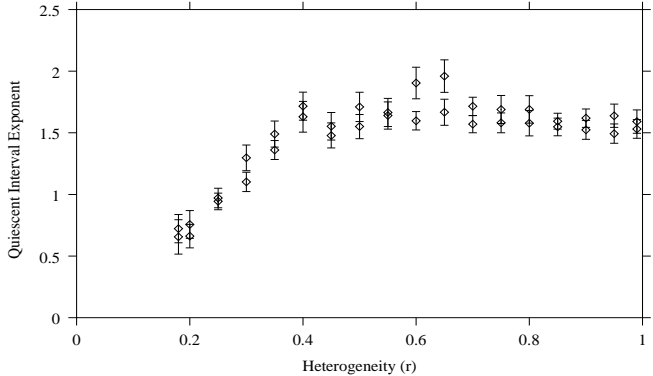, width=8.6cm,height=5cm}
\caption{Dependence of the exponent $x$ on fault
heterogeneity $r$ for a fault separation of 0.062 of the fault length.
The maximum likelihood exponent is plotted for both faults.}
\end{figure}

Tectonic loading plays no role
in the flipping of a fault from a dormant to
an active state since accumulating tectonic stress is continuously released
on the active fault. An active fault may flip
to quiescence shortly after the time that the dormant fault
receives a large enough force from
the active one to trigger one site on it into activity. We focus attention on
the fluctuations in force on the dormant fault.
For long fault lengths, the events that are remotely separated on a
single fault are
relatively independent. From the central-limit theorem, we expect the
distribution of fluctuations in stress on the active fault
to be gaussian and this is indeed the case. As the
annealing parameter $r$ increases, the increase in the range of rupture
thresholds translates to an increase in the variance of the
distribution of stresses; the larger the fluctuations in stress on the
active fault, the greater the chance of initiating rupture on the
inactive fault.
For small enough $r$, the activity on a
single fault will runaway
all too easily. For larger but also low
values of $r$, the variance of the stress on the active fault
is too small to have
significant
fluctuations of the stress field to trigger activity on the
quiescent fault. This leads to the following prediction: for two
faults with the same average threshold $B$  but with different $r$'s, we
expect the more
heterogeneous member, that is the one with the largest $r$, to produce
greater stress fluctuations on the active fault that might
activate the quiescent
fault. Therefore, the fault with the greater heterogeneity must be the
least active of the two.
As a test of this assertion, we find that only
fault $2$ is active in the case $r_1 = 75\%$ and $r_2 = 20\%$. Both
faults are active for $r_1 = 75\%$ and $r_2 = 55\%$, but fault $2$ is
active for more than 90\% of the time. As expected,
the exponent $x$ decreases with increasing
distance between the faults.

The amplitudes of these
fluctuations in stress depend on both the length and
position of the events on the active fault. Reactivation of the
quiescent fault will occur at the rare time that a large fracture
occurs on the active fault (but not so large as to cause runaway) so
that a large force will be generated near the edge of the crack;
the edge of this fracture has about the same $x$-coordinate as
the site with low quenched strength on the quiescent fault.
Thus, the quiescent fault will become active when a stress fluctuation
$\sigma_a$ induced by the active fault on it becomes larger than
$(B_q(x) - \sigma_q(x,t))_{min}$,
where $\sigma_q(x,t)$ is the stress quenched on the quiescent fault.
We use a mean field theory to describe the state of stress
$\sigma_a(x,t)$ due to an active fault on the opposing, dormant fault.
Suppose $\sigma_a$ fluctuates randomly due to fault activity with the
distribution
$D_{\sigma_a} = C {{e^{-{(\sigma_a - \sigma_0)^2}/2s_a^2}} \over
{\sqrt{2\pi}s_a}}$
for $0 \leq \sigma_a <  G(d) B(1+r)$.  $C$ is the normalizing factor which
accounts
for the compactness of the distribution support and $G(d)$ is the Green
function
that transfers stress from the active fault to the quiescent one.
$\sigma_0$ is the mean stress equal to $B G(d)$.
The standard deviation $s_a$ must
perforce be much less than $Br$.
There is a quenched stress `barrier'
$\Sigma=B_q(x)-\sigma_q(x)$ on the quiescent fault,
where $B_q$ and $\sigma_q$ are the frozen fracture thresholds and
stresses on it, the latter due to the nonuniform slip on it at the
time of onset of quiescence. We model the distribution of barriers $\Sigma$ as
$D_{\Sigma}(\Sigma)=C'{{e^{-{(\Sigma -\Sigma_0)^2}/2{s_\Sigma}^2}}
\over {\sqrt{2\pi}s_\Sigma}}$ where ${s_\Sigma}\ll Br$.
The variances $s_a^2$ and $s_{\Sigma}^2$ are in general different because
they reflect different stress fluctuations\,: $s_a^2$ is
averaged over all possible stress configurations induced by the active fault
on the dormant fault
while $s_{\Sigma}^2$ is controlled by the variability of the quenched
stress left over
from the last events before quiescence.
 
The distribution
of quiescent intervals $\Delta t$, for some fixed ${\Sigma}$, is the
same as the distribution of the number (proportional to $\Delta t$) of
events on
the active fault that are necessary
to get $\sigma$ larger than some fixed ${\Sigma}$:
$$
P_{{\Sigma}}(\Delta t) = \int_{\Sigma}^{\Sigma_{max}}dE'~D_{\sigma}(E')
\biggl( \int_{0}^{\Sigma}\!D_{\sigma}(E') dE' \biggl) ^{\Delta t -1}.
\eqno(2)
$$
The complete distribution of times to reactivation by an isolated event
on the quiescent fault is the sum of
(2) over all possible values of ${\Sigma}$ weighted by its
corresponding distribution:
$$
P(\Delta t) = \int_{0}^{\Sigma_{max}}~D_{\Sigma}({\Sigma})
~P_{\Sigma}(\Delta t)~d{\Sigma}~.
\eqno (3)
$$
For $\Delta t$ large, (2) and (3) give (1) with
$$
x = {s_a^2 \over {s_\Sigma}^2}~. \eqno(4)
$$
For the Weibull distribution $D_{\Sigma} = e^{-\left({\Sigma -B \over
s_{\Sigma}}\right)^n}$, and a similar expression for
$D_{\sigma}$ we get $x = s_a^n/{s_\Sigma}^n$. Very long tailed
power law distributions for $P(\Sigma)$ gives an exponential
distribution for (1).
 
Since $s_a$ is an increasing function of $r$,
we expect from (4) that
$x$ will also increase with $r$ for smaller values of $r$,
as observed (Fig.~3). For $r>0.4$, the
fluctuations in stress saturate
and $x$ becomes independent of $r$.
Also $s_a$ is a decreasing function of fault separation $(d)$ due to the
smoothing of stress inhomogeneities with distance.
Our results are robust whether the breaking strength distribution
is power law or uniform,
and the distributions need not be the same from fault to fault.
Our results are found to
hold for up to 10 faults in competition which is the maximum number of
faults we have considered.
 
The number of flips $m(t)$ during a time interval $t$ increases as
$m(t) \sim t^x$, and hence is a self-affine function of dimension $x < 1$
on the time
axis.  To see this, consider
$m$ successive events separated in time by $\Delta t_i, i=1,...,m ,$
where
$\Delta t_1+\Delta t_2+ ...+\Delta t_m = t = m \langle \Delta t \rangle$,
and
$$
\langle \Delta t \rangle \sim \int^{\Delta
t_{max}} dt {t \over t^{1+x}} \sim \Delta t_{max}^{1-x}~.
$$
 Since the maximum $\Delta t_{max}$ among $m$ trials is typically given by
$m \int_{\Delta t_{max}}^{\infty} {dt' \over t'^{1+x}} \sim 1$, we have $\Delta
t_{max} \sim m^{1 \over x}$. Thus $t = m
\langle \Delta t \rangle \sim m^{1 \over x}$, i.e. $m \sim t^x$,  which is
valid
for $x<1$. For $x>1$, $m \sim t$.
 
The power law distribution of time intervals between flips implies
nonstationarity and aging. Because of the self-similarity embodied in the
power-law distributions, we can state that the longer
since the last flip of activity, the longer the expected time till the
next \cite{Davis}. In other words, any expectation of a flip that is
estimated today depends on the past in a manner which does not decay. This
is a hallmark of aging. The mechanism is similar to the ``weak
breaking of ergodicity'' in spin glasses that occurs when the
exponent $x$ of the distribution of trapping times in metastable
states is less than one \cite{Bouchaud}.
 
Observations of
activity flipping between fault branches on a time
scale larger than inter-earthquake times are difficult to document due to the
scarcity of reliable historical data. In a well-documented case \cite{Bean},
localized flips in activity have been reported. Temporal clustering of
earthquakes on some faults has been identified (see \cite{CL90} for a
summary), but intermittency and fault interactions 
in the spirit of Fig. 1 have not been
observed, which would be needed
to establish the relevance of our flipping mechanism.
 
We have shown that the competition between faults leads to the persistence
phenomenon with a power law distribution of the durations of quiescent
phases. Our
results suggest that complexity in earthquakes and faulting may be a generic
outcome of the dynamics of interacting faults. We believe that a
broad study of paleoseismicity on many faults is needed
to identify intermittency as we have described it.

{\bf Acknowledgements:}
We are grateful to C. Vanneste for help with programming and to the late
S.Beanland for calling her work to our attention.
Publication no. XXXX Institute of Geophysics and
Planetary Physics, University of California, Los Angeles.

\end{document}